\documentstyle[subfigure,epsfig,amsmath,amssymb,]{elsart}
\begin{document}
\begin{frontmatter}
\title{Shell structure in the density profile of a rotating gas of
spin-polarized fermions}
\author{Z. Akdeniz$^{1,2}$}, 
\author{P. Vignolo$^{1}$\corauthref{cor1}} and 
\author{M.P. Tosi$^1$}
\corauth[cor1]{Corresponding author, e-mail: {\tt vignolo@sns.it}}
\address{$^1$NEST-INFM and
Classe di Scienze, Scuola Normale Superiore, I-56126 Pisa, Italy\\
$^2$ Physics Department, Istanbul University, Istanbul, Turkey}
\maketitle
\begin{abstract}
We present analytical expressions and numerical illustrations 
for the ground-state density distribution of an ideal gas of spin-polarized 
fermions moving in two dimensions and driven to rotate in a harmonic 
well of circular or elliptical shape. We show that with suitable 
choices of the strength of the Lorentz force for charged fermions, 
or of the rotational frequency for neutral fermions, 
the density of states can be tuned as a function of the angular 
momentum so as to display a prominent shell structure in the 
spatial density profile of the gas. We also show how this feature 
of the density profile is revealed in the static structure factor 
determining the elastic light scattering spectrum of the gas.
\end{abstract}


\end{frontmatter}

\section{Introduction}
Ultracold Fermi gases of alkali atoms such as $^{40}$K or $^6$Li in harmonic 
traps are quantum systems
that are experimentally accessible by the techniques of atom 
trapping and cooling \cite{ref8,ref9}.
The high purity and the low temperature
of the samples and the high resolution of the
detection techniques make these systems ideal candidates for the study
on the mesoscopic scale of single-level quantum properties such as
shell structures in the particle 
density profiles \cite{refA,refB,refC,refD,refG}. 
In the experiments the trapped atomic gas can be fully spin-polarized and
the strength and the
anisotropy of the trap can be tuned to reach 
quasi-onedimensional (1D) or
quasi-twodimensional (2D) configurations. 
Quantum effects in the equilibrium profiles can be greatly enhanced
by varying the anisotropy of the confinement \cite{refD}.
Here we show how to enhance quantum effects on 2D equilibrium density
profiles by putting the gas into rotation.

In a wholly different physical realm electrons in a quantum dot realize
a Fermi gas moving 
in a plane under the effect of a harmonic potential.
A quantum dot is often described as an 
artificial atom, in which a confinement potential replaces the 
attractive potential of the nucleus 
and the increased length scale
allows access to experiments that are not practicable on real 
atoms \cite{ref1,ref2,ref3}.
Spin effects are also accessible 
and complete spin polarization is easily achieved in a small 
dot.
 Confinement by a circularly symmetric harmonic potential is of 
special interest, since in this case the energy needed to add 
electrons to the dot reveals a shell structure \cite{ref4}. The effects 
of a magnetic field on the electronic properties of circular 
quantum dots have been studied by numerical diagonalization 
and by spin-density functional methods (see Ref. \cite{ref5} and references 
therein). These approaches have also been applied to rectangular 
dot structures, when the effective lateral confinement is described 
as elliptical. In such a confinement the single-particle energy 
levels as well as the energy spectra for two-electron systems 
are exactly known as functions of the magnetic field \cite{ref6,ref7}.

A Fermi gas of spin-polarized charged 
particles in a uniform magnetic field, under conditions such that
the Coulomb interactions 
can be neglected as specified below, can be mapped into
a rotating Fermi gas of neutral atomic particles in a state of complete 
spin polarization, where the atom-atom interactions are negligible 
on account of the Pauli principle suppressing $s$-wave scattering. 
 In this work we examine the particle density profiles of the ideal 
system that we have just introduced. For definiteness we shall 
refer throughout the paper to the ideal 2D Fermi gas of charged 
particles in a uniform magnetic field, although some parts of 
the discussion (and in particular the analysis of the elastic 
light scattering spectrum given in Sec. \ref{seclight}) are specifically 
aimed at an atomic Fermi gas in uniform rotation.

 An ideal 2D Fermi gas of charged particles in a uniform magnetic 
field has recently been studied by van Zyl and Hutchinson \cite{ref10}, 
who have used an inverse Laplace transform method and an expansion 
in Laguerre polynomials to calculate its thermodynamic properties. 
Their method can be implemented exactly in the case of a uniform 
gas, whereas recourse to a local-density approximation is needed 
for an inhomogeneous gas on account of the broken symmetry induced 
by the concurrent effects of the harmonic confinement and of 
the magnetic field. However, the single-particle Hamiltonian 
of the ideal gas can be diagonalized on the basis of left and 
right circular or elliptical quanta, bringing it to the form 
of the Hamiltonian of two independent harmonic oscillators. The 
eigenfunctions can thus be written in terms of Hermite polynomials, 
as we do in this work. Practical applications of our approach 
are limited to ultracold Fermi gases consisting of a restricted 
number of particles, so that the number of Hermite polynomials 
that need explicit numerical calculation remains limited. We 
shall give below numerical illustrations for systems of 10 and 
106 fermions at zero temperature.

 The paper is organized as follows. In Sec. \ref{sec2} we set out the 
single-particle Hamiltonian of the system in a generic gauge 
which allows its diagonalization on the basis of left and right-handed 
circular or elliptical quanta. In Sec. \ref{sec3} we express the spatial 
density profile of the gas in terms of Hermite polynomials and 
display several configurations for various values of the strength 
of the magnetic 
field and of the anisotropy parameters of the confinement. We show that 
an impressive shell structure arises in the profile whenever there 
is an excess of density of states with zero angular momentum. 
In Sec. \ref{seclight} we calculate the elastic contribution to the static 
structure factor of the gas, in order to show how the shell structure 
is reflected in the elastic light scattering spectrum. Finally 
Sec. \ref{sec5} offers some concluding remarks.

\section{The model}
\label{sec2}

 We consider a 2D spin-polarized Fermi gas at zero temperature, 
made up of $N$ non-interacting fermions of charge $e$ and 
mass $m_f$ subject to a uniform magnetic field 
${\bf B} =\boldsymbol{\nabla}\times{\bf A}$  perpendicular 
to the plane of confinement (the $\{x,y\}$ plane, say). 
The gas is immersed in a uniformly charged background medium 
so that the total system is neutral. Screening effects cut the 
range of the Coulomb interactions and to a first approximation 
these can be neglected on account of the Pauli principle keeping 
fermions with parallel spins far apart from each other. 
More generally, the Coulomb interactions between electrons in a quantum dot
are negligible when the harmonic oscillator length characterizing the 
confinement is small compared with the effective Bohr radius.

As already 
noted in Sec. I, this system can be mapped into a spin-polarized 
gas of $N$ neutral fermions confined in a rotating trap. The map
is guaranteed by a
correspondence between the Maxwell equations and
the Navier-Stokes equations in the absence of viscosity, 
as demonstrated by Marmanis \cite{ref10b}.
In the rotating system the velocity field ${\bf v}$ of the fermions
and the vorticity $\boldsymbol{\omega}=\boldsymbol{\nabla}\times{\bf v}$
play the role of the vector potential ${\bf A}$ and of
the magnetic field ${\bf B}$, respectively.

\subsection{Circular trap}
 It is useful to present first the case of an isotropic harmonic 
potential. The single-particle Hamiltonian for non-interacting 
fermions is
\begin{equation}
\hat H=\dfrac{1}{2m_f}(\hat p_x^2+\hat p_y^2)
+\dfrac{1}{2}m_f[(\omega_0^2+\omega_L^2)(\hat x^2+\hat y^2)]
+\omega_L(\hat x\hat p_y-\hat p_x\hat y)\,.
\label{eq1}
\end{equation}
In Eq. (\ref{eq1}) ${\omega}_{0}$ is the trap frequency, 
$\omega_L=eB/(2cm_f)$ is the Larmor 
frequency, and we have chosen the symmetric gauge for the vector 
potential ${\bf A} = (-{By}/2, {Bx}/2, 0)$, as is suited to the 
circular symmetry of the confinement \cite{ref10t}.

 The Hamiltonian (\ref{eq1}) is separable \cite{ref11} upon introducing first 
the creation and destruction operators $\hat a_{x,y}^\dagger$ and
$\hat a_{x,y}$ for a quantum of oscillation 
in the $x$ or $y$ direction, defined according to 
$\hat a_{x}=[(\hbar/m_f\Omega)^{-1/2}\hat x+
i(\hbar m_f\Omega)^{-1/2}\hat p_x]/\sqrt{2}$ where 
$\Omega=(\omega_0^2+\omega_L^2)^{1/2}$, and then 
the operators $\hat a_d=(\hat a_x-i\hat a_y)/\sqrt{2}$ and
$\hat a_g=(\hat a_x+i\hat a_y)/\sqrt{2}$. The transformed Hamiltonian is
\begin{equation}
\hat H=\hbar\omega_d\left(\hat a_d^\dagger\hat a_d+\dfrac{1}{2}\right)+
\hbar\omega_g\left(\hat a_g^\dagger\hat a_g+\dfrac{1}{2}\right)
\label{eq2}
\end{equation}
where $\omega_d=\Omega+\omega_L$ and $\omega_g=\Omega-\omega_L$.

 The operator  $\hat a_d$ ($\hat a_g$) can be interpreted as the 
destruction operator 
of a right (left) circular quantum, the number of such quanta 
being given by $\hat n_d=\hat a_d^\dagger \hat a_d$ 
($\hat n_g=\hat a_g^\dagger \hat a_g$). 
The quantity $\hbar(\hat n_d-\hat n_g)$ gives the eigenvalues of the 
angular momentum operator $\hat L_z=\hat x\hat p_y-\hat y\hat p_x$.

\subsection{Elliptical trap}
 In the case of anisotropic 2D confinement neither the symmetric 
gauge nor the Landau gauge are useful. We write the single-particle 
Hamiltonian in the general form
\begin{equation}
\hat H=\dfrac{1}{2m_f}(\hat p_x-\alpha eB \hat y/c)^2+
\dfrac{1}{2m_f}(\hat p_y+(1-\alpha)\,eB \hat x/c)^2
+\dfrac{1}{2}m_f\omega_0^2(\hat x^2+\lambda^2\hat y^2)
\label{eq3}
\end{equation}
with $0 <\alpha< 1$ and $\lambda$ the anisotropy 
parameter. Equation (\ref{eq3}) can be rewritten as
\begin{equation}
\hat H=\dfrac{1}{2m_f}(\hat p_x^2+\hat p_y^2)
+\dfrac{1}{2}m_f(\omega_x^2 \hat x^2+\omega_y^2\hat y^2)
+2\omega_L\left[(1-\alpha)\hat x\hat p_y-\alpha \hat p_x\hat y\right]
\label{eq4}
\end{equation}
where we have set $\omega_x^2=\omega_0^2+4(1-\alpha)^2\omega_L^2$ and
$\omega_y^2=\lambda^2\omega_0^2+
4\alpha^2\omega_L^2$. 
By imposing the condition 
$(1-\alpha)/{\alpha}={\omega_x(\alpha)}/{\omega_y(\alpha)}$, which has 
the solution
\begin{equation}
\alpha=\lambda/(1+\lambda)\,,
\label{eq5}
\end{equation}
and using $\hat a_{x}=[(\hbar/m_f\omega_x)^{-1/2}\hat x+
i(\hbar m_f\omega_x)^{-1/2}\hat p_x]/\sqrt{2}$ etcetera, 
Eq. (\ref{eq4}) becomes
\begin{eqnarray}
\hat H&=&\dfrac{1}{2}\hbar\left(\omega_x+
\omega_y\right)(\hat a_x^\dagger\hat a_x+
\hat a_y^\dagger\hat a_y +1) +
\dfrac{1}{2}\hbar\left(\omega_x-\omega_y\right)(\hat a_x^\dagger\hat a_x-
\hat a_y^\dagger\hat a_y)\nonumber\\
&+&2\dfrac{\sqrt{\lambda}}{1+\lambda}
i\hbar\omega_L\left[
\hat a_x\hat a_y^\dagger-\hat a_x^\dagger\hat a_y\right]\,.
\label{eq6}
\end{eqnarray}
For circular confinement ($\lambda = 1$) the $\alpha$-gauge 
given by the relation in Eq. (5) reduces to the symmetric gauge, 
while in the limit of one-dimensional confinement ($\lambda 
= 0$ or $\infty$) it reduces to the Landau gauge.
More generally, in the mapping with a rotating system
the $\alpha$-gauge corresponds to a velocity field which follows
the symmetry of the confinement.

 The Hamiltonian in Eq. (6) is separable in terms of the operators
\begin{equation}
\left\{\begin{array}{l}
\hat a_d=\cos\theta\,\hat a_x-i \sin\theta\, \hat a_y\\
\hat a_g=\sin\theta\, \hat a_x+i\cos\theta\, \hat a_y\\
\end{array}\right.
\label{eq7}
\end{equation}
if the angle $\theta$ satisfies the relation 
$\tan(2\theta)=2\tilde\omega_L/(\omega_x-\omega_y)$, with
$\tilde\omega_L=2\sqrt{\lambda}\omega_L/(1+\lambda)$. In 
the new basis the Hamiltonian can be written as in Eq. (\ref{eq2}) with
\begin{equation}
\omega_{d,g}=\dfrac{1}{2}(\omega_x+\omega_y)
\pm\sqrt{\tilde\omega_L^2+\dfrac{1}{4}(\omega_x-\omega_y)^2}
\label{eq8}
\end{equation}
The operators $\hat a_d$ and $\hat a_g$ are the destruction 
operators for a right-handed 
and a left-handed elliptical quantum, but (if $\theta\neq\pi/4$) the 
quantum number $m=n_d-n_g$ is no longer proportional to the angular 
momentum $L_z$.

\section{Spatial density profiles}
\label{sec3}
 The eigenstate $|\chi_{n_d,n_g}\rangle$
corresponding to the occupation numbers $\{n_d,n_g\}$ 
can be obtained by recursively applying the operators 
$\hat a^\dagger_d$ and $\hat a^\dagger_g$ to 
the ground state $|\chi_{0,0}\rangle$. Using Eq. (\ref{eq7}) we can write
\begin{equation}
|\chi_{n_d,n_g}\rangle=\dfrac{1}{\sqrt{n_d!n_g!}}
(\cos\theta \hat a_x^\dagger+i\sin\theta \hat a_y^\dagger)^{n_d}
(\sin\theta \hat a_x^\dagger-i\cos\theta \hat a_y^\dagger)^{n_g}
|\chi_{0,0}\rangle\,,
\label{eq9}
\end{equation}
where
\begin{equation}
\langle x,y|\chi_{0,0}\rangle=\pi^{-1/2}
\exp(-x^2/2l_x^2)\exp(-y^2/2l_y^2)
\label{eq10}
\end{equation}
with $l_{x,y}=\sqrt{\hbar/m_f\omega_{x,y}}$. 
Series expansion of Eq. (\ref{eq9}) and use of the expression 
of the eigenstates of the harmonic oscillator in terms of Hermite 
polynomials lead to the result
\begin{equation}
\begin{split}
\chi_{n_d,n_g}(x,y)&=\dfrac{\exp(-x^2/2l_x^2)\exp(-y^2/2l_y^2)}
{\sqrt{\pi\,l_xl_y\, 2^{n_d+n_g}}}
\sum_{j=0}^{n_d}\sum_{k=0}^{n_g}
\dfrac{\sqrt{n_d!\,n_g!}}{j!(n_d-j)!k!(n_g-k)!}(-1)^{n_g-k}\\
&(i)^{n_d+n_g-k-j}
(\sin\theta)^{n_d-j+k}
(\cos\theta)^{n_g-k+j}
H_{j+k}(x/l_x)H_{n_d+n_g-j-k}(y/l_y).\\
\end{split}
\label{eq11}
\end{equation}
Finally, the density profile $n(x,y)$ can be expressed 
in the form
\begin{equation}
n(x,y)=\sum_{n_g=0}^{N_g}\sum_{n_d=0}^{N_d}
|\chi_{n_d,n_g}(x,y)|^2\,.
\label{eq12}
\end{equation}
Here,
\begin{equation}
N_d={\rm Int}
\left[\frac{E_F}{\hbar\omega_d}-
\left(n_g+\dfrac{1}{2}\right)\dfrac{\omega_g}{\omega_d}
-\dfrac{1}{2}\right]\,
\label{eq13}
\end{equation}
is the highest allowed value for the quantum number $n_{d}$ at 
given values of the Fermi energy $E_{F}$ and of the quantum number 
$n_{g}$, and
\begin{equation}
N_g={\rm Int}
\left[\frac{E_F}{\hbar\omega_g}-
\dfrac{\omega_d}{2\omega_g}-\dfrac{1}{2}\right]
\label{eq14}
\end{equation}
is the highest allowed value for the quantum number $n_{g}$ at 
given $E_{F}$. Of course, the Fermi energy is determined by the 
number of fermions in the trap.

\begin{figure}
\centering{
\epsfig{file=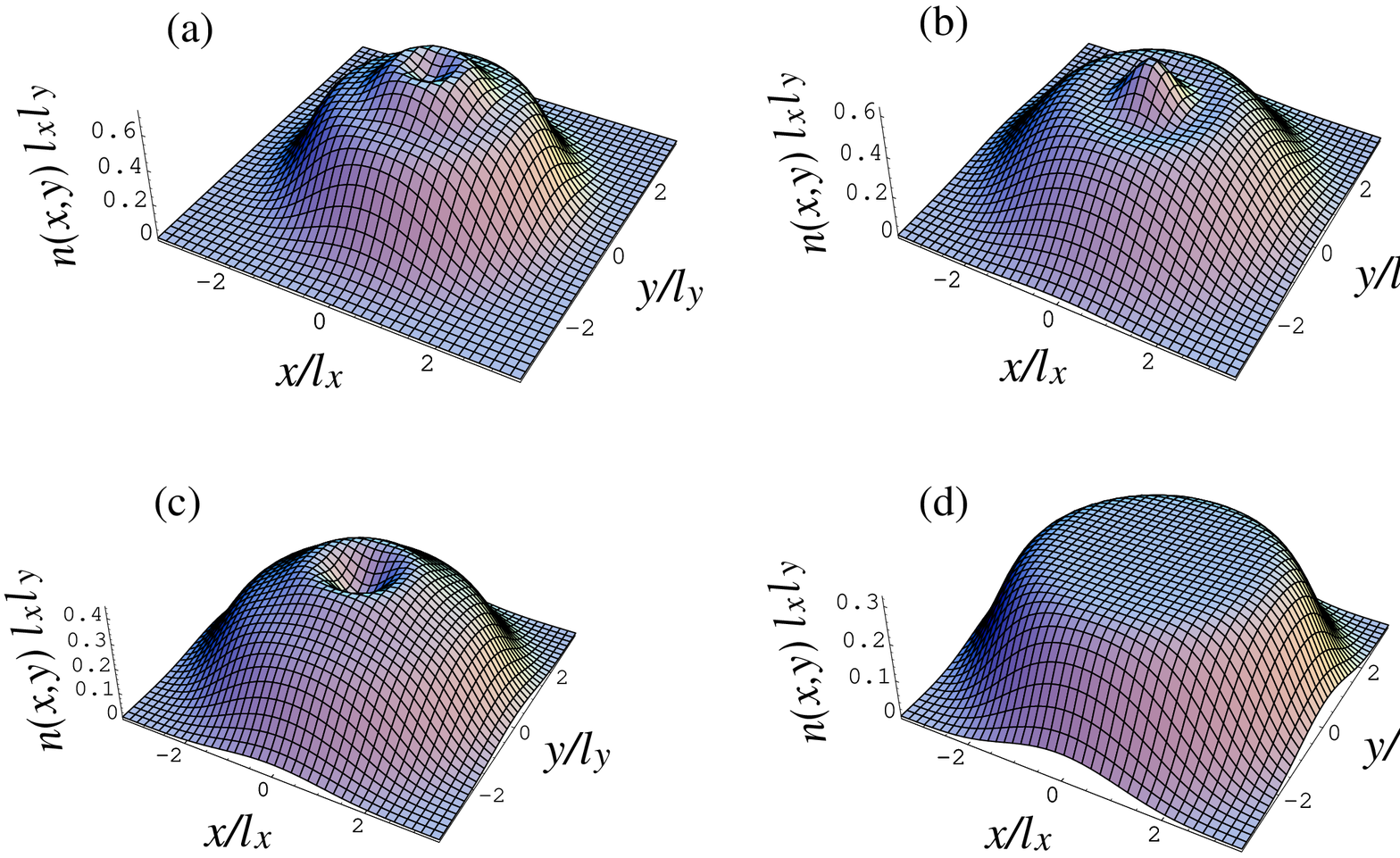,width=0.7\linewidth}}
\caption{Density profile for 10 fermions in a circular harmonic 
trap at (a) $\omega_{L}$ = 0, (b) $\omega_{L}$ = $\omega_{0}$, 
(c) $\omega_{L}$ = 1.18$\omega_{0}$, and (d) $\omega_{L}$ 
= 10$\omega_{0}$.}
\label{fig1}
\end{figure}

\begin{figure}
\centering{
\epsfig{file=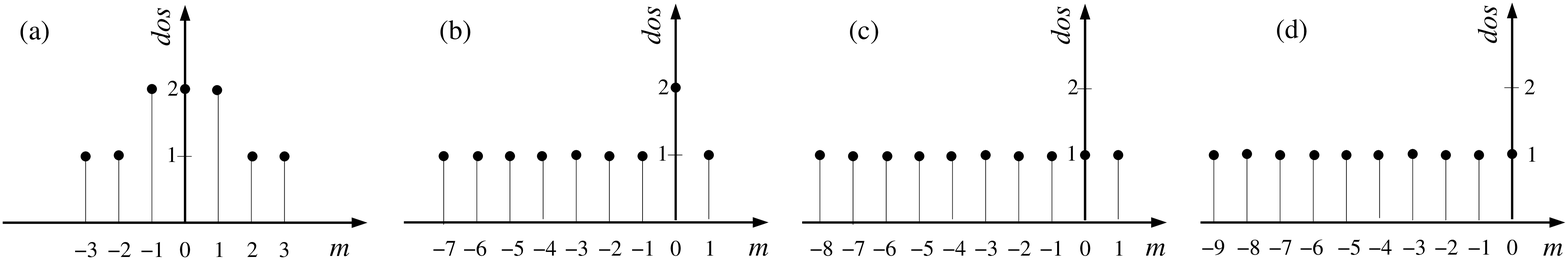,width=1.\linewidth}}
\caption{Density of occupied states (dos) as a function 
of the quantum number \textit{M} = \textit{n}$_{d}$ -- \textit{n}$_{g}$ for each of the 
configurations shown in Fig. \ref{fig1}.}
\label{fig2}
\end{figure}

 Figure \ref{fig1} shows the particle density profile for 10 spin-polarized 
fermions in an isotropic trap ($\lambda = 1$) at various 
strengths of the applied magnetic field, and Fig. \ref{fig2} reports the 
corresponding density of occupied states (dos). In zero field 
(Figs. \ref{fig1}(a) and \ref{fig2}(a)) the total angular momentum is zero and 
the left and right states are equally populated. At a field strength 
such that the Larmor frequency is equal to the trap frequency 
(Figs. \ref{fig1}(b) and \ref{fig2}(b)) there still is an excess population of 
the state $m= 0$ and a central peak in the density profile. 
At still higher fields (Figs. \ref{fig1}(c) and \ref{fig2}(c)) the dos becomes 
constant and the occupation of two states with opposite angular 
momentum ($m= \pm 1$) produces a central hole in the 
density profile. Finally, at very large fields (Figs. \ref{fig1}(d) and 
\ref{fig2}(d)) the Landau quantization limit is being recovered, with 
a filling factor equal to unity in a dos which 
is constant for $m\le0$. This is spatially the most compact 
distribution of the droplet of spin-polarized fermions, which 
is commonly denoted as MDD \cite{ref12}.

 Similar density distributions can be obtained for higher numbers 
of fermions. In Fig. \ref{fig3} we report as an example the spatial 
distribution 
of 106 fermions at $\omega_{L}= 5\omega_{0}$, showing 
as in Fig. \ref{fig1}(b) a central peak from an excess dos at $m=0$. 
On a slight increase in the field strength, to $\omega_{L}
\ge 5.1\omega_{0}$, the filling factor becomes unity and the 
density profile takes a MDD configuration similar to that shown 
in Fig. \ref{fig1}(d).

 In an anisotropic trap the overall elliptical shape of the fermion 
cloud can still sustain a shell structure, which becomes progressively 
smoothen as the field is increased. In Fig. \ref{fig4} we show the 
case of 10 fermions in a trap with anisotropy parameter $\lambda 
= 4$. For $\omega_{L} = \omega_{0}$ (Fig. \ref{fig4}(b)) the 
dos is similar to that shown for a circular trap in Fig. \ref{fig2}(b), 
but the value of $m$ is not related to the angular momentum 
in the anisotropic trap and a central peak in the density profile 
is consequently missing. Instead, raising the field produces 
a softening of the oscillations of the profile along the \textit{x} 
axis and some enhancement of those along the $y$ axis. This 
effect is more pronounced at $\omega_{L} = 1.8\omega_{0}$ 
(Fig. \ref{fig4}(c)), where a central hole in the profile is accompanied 
by two bumps along the \textit{y} axis. For this value of $\omega_L$
the dos is similar to that shown for a circular trap in Fig. \ref{fig2}(c).
Again, all shell structures 
disappear and an MDD-like profile is obtained when the filling 
factor becomes unity (Fig. \ref{fig4}(d)).

 Finally, Fig. \ref{fig5} reports the density profile for 10 fermions in 
a strongly anisotropic trap ($\lambda = 10$). In zero field 
(Fig. \ref{fig5}(a)) the system is quasi-onedimensional: only the ground 
state is occupied in the $y$ direction and density oscillations 
are present only along the $x$ axis. These become progressively 
smoothen as the field is applied (Fig. \ref{fig5}(b)) and increased 
(Fig. \ref{fig5}(c)).

\section{Angular distribution of scattered light}
\label{seclight}

 The theory of light scattering from a confined cloud of spin-polarized 
fermionic atoms has been presented in Ref. \cite{ref13}. Here we discuss 
an experiment of elastic scattering in which the incident beam 
is propagating along the $z$ direction, orthogonally to the 
$\{x,y\}$ plane of confinement of the gas.

 The positive-frequency component of the incident electric field 
is
\begin{equation}
{\bf E}_F^+(z,t)= \frac{1}{2} E \hat \varepsilon \exp[i(k_Lz-\omega t)]
\label{eq15}
\end{equation}
with $E$ and $\hat\varepsilon$ being the field amplitude and polarization vector, 
and ${k}_{L}$ and $\omega$ its wave number and frequency. 
The angular distribution of the scattered light can be decomposed 
into two components: an elastic term originating from diffraction 
by a finite object of given optical density profile, and an inelastic 
term determined by excitations in the quantum fluid. Only the 
elastic term remains if the light frequency is far-off-resonance 
from the internal states of the fluid. This term is determined 
by the elastic static structure factor $S_e(k_x,k_y)$, which is defined by
\begin{equation}
S_e(k_x,k_y) = \left|\int dx\int dy \, e^{-i k_x
x}e^{-i k_yy} n(x,y) \right|^2
\label{eq16}
\end{equation}
where ${\bf k} = ({k}_{x},{k}_{y})$ is the wave vector transfer 
in the scattering process.

\begin{figure}
\centering{
\epsfig{file=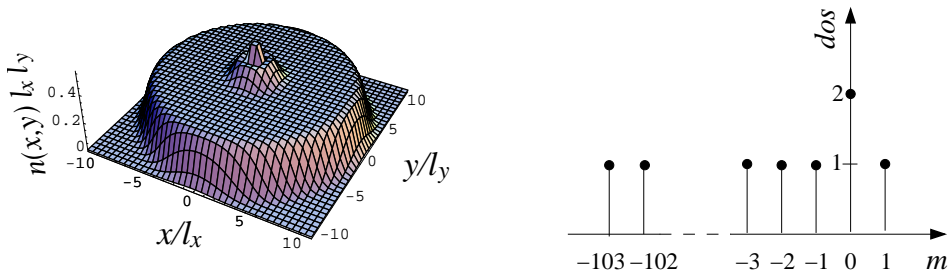,width=0.9\linewidth}}
\vspace{0.2cm}
\caption{Density profile (left) and dos (right) for 106 
fermions in a circular harmonic trap at $\omega_{L}$ = 5$\omega_{0}$.}
\label{fig3}
\end{figure}

\begin{figure}
\centering{
\epsfig{file=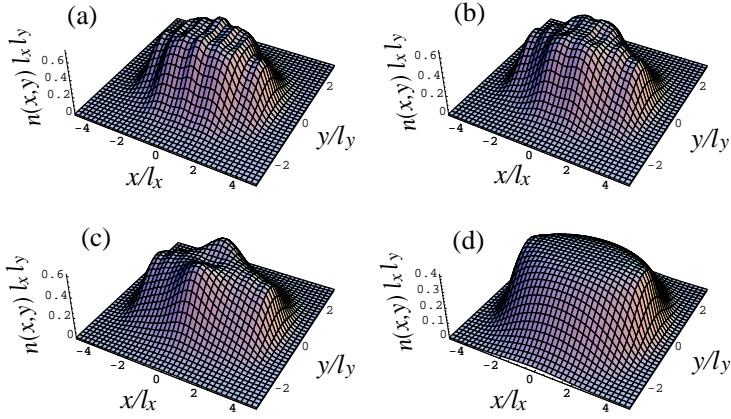,width=0.7\linewidth}}
\caption{Density profile for 10 fermions in an elliptical 
harmonic trap with anisotropy parameter $\lambda$ = 4 at 
(a) $\omega_{L}$ = 0, (b) $\omega_{L}$ = $\omega_{0}$, 
(c) $\omega_{L}$ = 1.8$\omega_{0}$, and (d) $\omega_{L}$ 
= 10$\omega_{0}$.}
\label{fig4}
\end{figure}

\begin{figure}
\centering{
\epsfig{file=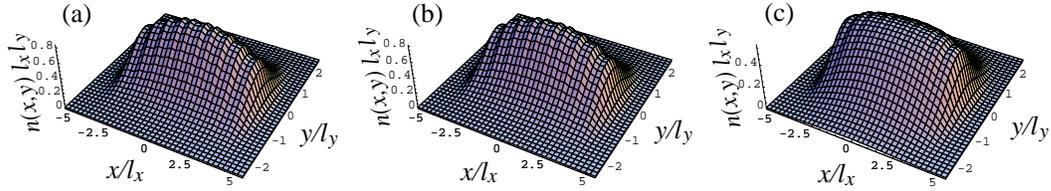,width=1\linewidth}}
\caption{Density profile for 10 fermions in an elliptical 
harmonic trap with strong anisotropy ($\lambda$ = 10) at 
(a) $\omega_{L}$ = 0, (b) $\omega_{L}$ = $\omega_{0}$, 
and (c) $\omega_{L}$ = 10$\omega_{0}$.}
\label{fig5}
\end{figure}

\begin{figure}
\centering{
\epsfig{file=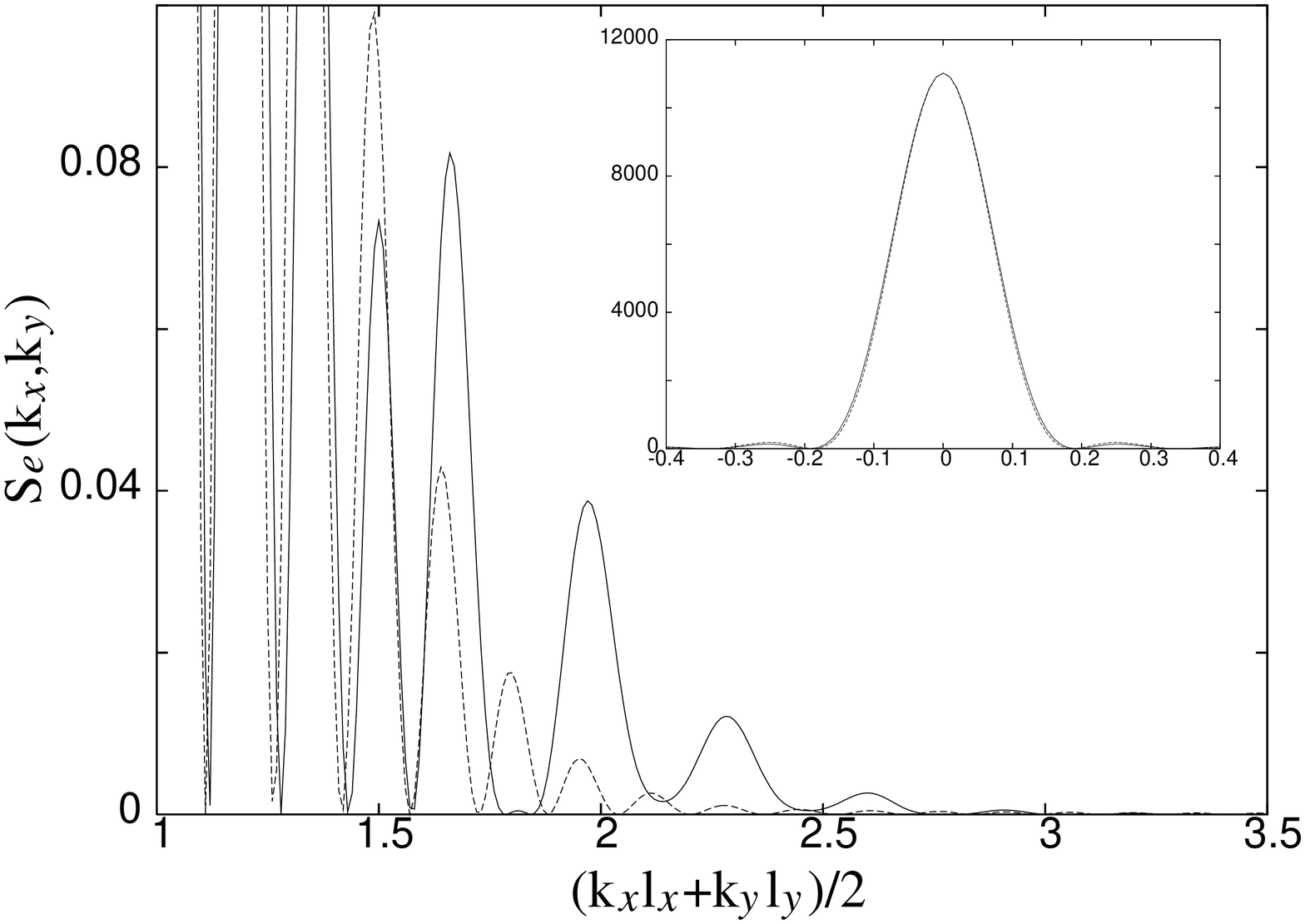,width=0.6\linewidth}}
\caption{Elastic contribution to the static structure factor $S_e(k_x,k_y)$ 
corresponding to the spatial density distribution shown in Fig. 
\ref{fig3} (continuous line) and to an MDD distribution of the same number 
of fermions (dashed line), as a function of  $(k_xl_x+k_yl_y)/2$ at $k_xl_x=k_yl_y$. 
The inset shows the central peak of the two patterns.}
\label{fig6}
\end{figure}

 We have evaluated the elastic contribution to the static structure 
factor for the density distribution of 106 fermions shown in 
Fig. \ref{fig3} and for a MDD configuration found with the same number 
of fermions at $\omega_{L}= 5.1\omega_{0}$. These 
two configurations have essentially the same size and thus show 
almost identical central peaks in their diffraction pattern (see 
the inset in Fig. \ref{fig6}). According to Babinet's principle this peak 
is given by the optical diffraction pattern of a circular aperture.

 The central part of the density profile shown in Fig. \ref{fig3} consists 
of a peak of diameter $\sim 2l_{x}$ surrounded by a ring of diameter 
$\sim 4l_{x}$. As noted in Sec. \ref{sec3}, this double structure arises 
from an excess of occupied states with zero angular momentum 
and its signal appears in the diffraction pattern in the range 
$1.5 <k_{x}l_{x}< 3$. In the same range the scattered 
intensity from the flat MDD configuration is much lower, as is 
seen in the main frame of Fig. \ref{fig6}.

\section{Summary and conclusions}
\label{sec5}

 In summary, we have evaluated the particle density profiles of 
an ideal gas of spin-polarized fermions confined in 2D inside 
a circular or elliptical harmonic trap as functions of an applied magnetic 
field or of the trap rotational frequency. We have introduced 
a generic $\alpha$-gauge in which the eigenstates of the 
Hamiltonian can be expressed in terms of right-handed and left-handed 
elliptical quanta and written in terms of Hermite polynomials. 
The $\alpha$-gauge reduces to the symmetric gauge in a 
circular trap and to the Landau gauge in a strongly elongated 
trap.

 In the case of a circular trap we have correlated the shell structures 
in the density profile of the gas with the density of occupied 
angular-momentum states and suggested that these structures may 
become observable in an elastic light scattering experiment from 
an atomic Fermi gas inside a rotating pancake-shaped trap. Other 
peculiar shell structures have been illustrated in the case of 
an elliptical trap. It remains a challenge to provide an exact 
treatment for a gas consisting of very large numbers of particles 
at finite temperature.

\ack
 This work was partially supported by MIUR through the PRIN2003 
program and by INFM through the PRA-Photonmatter Initiative. 
Z.A. acknowledges support from TUBITAK and from
the Research Fund of Istanbul University under
Project Number Y\"OP-16/13082004.

\newpage


\vspace{3cm}

\begin{thebibliography}{100}
 \bibitem{ref8} B. DeMarco and D.S. Jin, Science 285 (1999) 1703; 
M.O. Mewes, G. Ferrari, F. Schreck, A. Sinatra, and C. Salomon, Phys. 
Rev. A 61 (2000) 011403; C.A. Regal, M. Greiner, and D.S. 
 Jin, Phys. Rev. Lett. 92 (2004) 040403; L. Pezz\`{e}, L. 
Pitaevskii, A. Smerzi, S. Stringari, G. Modugno, E. de Mirandes, 
F. Ferlaino, H. Ott, G. Roati, and M. Inguscio, Phys. Rev. Lett. 93 
(2004) 120401.

 \bibitem{ref9} For a recent review on confined quantum gases see A. Minguzzi, 
S. Succi, F. Toschi, M.P. Tosi, and P. Vignolo, Phys. Rep. 395 (2004) 
223.

\bibitem{refA}
P. Vignolo, A. Minguzzi, and M.P. Tosi, Phys. Rev. Lett. 85 (2000) 2850.
\bibitem{refB}
F. Gleisberg, W. Wonneberger, U. Schl\"oder, and C. Zimmermann,
Phys. Rev. A 62 (2000) 063602.

\bibitem{refC}
M. Brack and B.P. van Zyl, Phys. Rev. Lett. 86 (2001) 1574.

\bibitem{refD}
P. Vignolo and A. Minguzzi, Phys. Rev. A 67 (2003) 053601.

\bibitem{refG}
E.J. Mueller, Phys. Rev. Lett. 93 (2004) 190404.

\bibitem{ref1} M.A. Kastner, Phys. Today 46 (1993) 24.

 \bibitem{ref2}T. Chakraborty, \textit{Quantum Dots: A Survey of the Properties 
of Artificial Atoms} (Elsevier, Amsterdam, 1999).

 \bibitem{ref3} L. Jacak, P. Hawrylak, and A. W\`{o}js, \textit{Quantum Dots} (Springer, 
Berlin, 1998).

 \bibitem{ref4} S. Tarucha, D.G. Austing, T. Honda, R.J. van der Hage, and 
L.P. Kouwenhoven, Phys. Rev. Lett. 77 (1996) 3613.

 \bibitem{ref5} S.M. Reimann and M. Manninen, Rev. Mod. Phys. 74 (2002) 1283.

 \bibitem{ref6} A.V. Madhav and T. Chakraborty, Phys. Rev. B 49 (1994) 8163.

 \bibitem{ref7} P.S. Drouvelis, P. Semelcher, and F.K. Diakonos, J. Phys.: 
Condens. Matter 16 (2004) 3633.


\bibitem{ref10} B.P. van Zyl and D.A.W. Hutchinson, Phys. Rev. B 69 (2004) 
024520.

\bibitem{ref10b} H. Marmanis, Phys. Fluids 10 (1998) 1428;
{\it ibid.} 3031.

\bibitem{ref10t}
Z.F. Ezawa, {\it Quantum Hall Effects: Field Theoretical Approach and Related
Topics} (World Scientific, Singapore, 2000).

\bibitem{ref11} C. Cohen-Tannoudji, B. Diu, and F. Lalo\"{e}, 
\textit{Mechanique Quantique} (Hermann, Paris, 1998).

\bibitem{ref12} S. Tarucha, D.G. Austing, S. Sasaki, Y. Tokura, W. van der 
Wiel, and L.P. Kouwenhoven, Appl. Phys. A 71 (2000) 367.

 \bibitem{ref13} P. Vignolo, A. Minguzzi, and M.P. Tosi, 
Phys. Rev. A 64 (2001) 023421.

\end{thebibliography}
\end{document}